\documentclass[twocolumn,secnumarabic,amssymb, nobibnotes, aps, prd,nofootinbib,fleqn, floatfix,showpacs]{revtex4-1}
\usepackage[utf8]{inputenc}
\usepackage{graphicx} 
\usepackage{amsmath,amssymb}
\usepackage{multirow}

\begin{document}

\title{Density dependence of the pairing interaction and pairing correlation in unstable nuclei}%

\author{S. A. Changizi}%
\email{asiyeh@kth.se}
\author{C. Qi}%
\email{chongq@kth.se}
\affiliation{Department of Physics, Royal Institute of Technology (KTH), SE-10691 Stockholm, Sweden}
\date{\today}%

\begin{abstract}
This work aims at a global assessment of the effect of the density dependence of the zero-range pairing interaction. Systematic
Skyrme-Hartree-Fock-Bogoliubov calculations with the volume, surface and mixed pairing forces are carried out to study the pairing gaps in even-even nuclei over the whole nuclear chart. Calculations are also done in coordinate representation for unstable semi-magic even-even nuclei. The calculated pairing gaps are compared with empirical values from four different odd-even staggering formulae.
Calculations with the three pairing interactions are comparable for most nuclei close to $\beta$-stability line. However, the surface interaction calculations predict neutron pairing gaps in neutron-rich nuclei that are significantly stronger than those given by the mixed and volume pairing. On the other hand, calculations with volume and mixed pairing forces show noticeable reduction of neutron pairing gaps in nuclei far from the stability. 
\end{abstract}

\maketitle

\section{Introduction}

The odd-even staggering (OES) of nuclear binding energy implies that the masses of odd nuclei are larger than the two adjacent even nuclei and pairing correlation has been associated with this effect \cite{PhysRev.10.936,bohr1998nuclear}. Pairing is a kind of emergent phenomenon underlying many aspects of the dynamics of atomic nuclei and is the most crucial correlation beyond the nuclear mean field. Of particular interest nowadays is the study of pairing  correlation properties in dripline nuclei where the pairing gap energy becomes comparable to the nucleon separation energy and the continuum effect may manifest itself. It turns out that the Hartree-Fock-bogoliubov (HFB) approach with effective zero-range pairing forces is a reliable and computational convenient way to study the nuclear pairing correlations in both of both stable and unstable nuclei (see, e.g., Refs. \cite{Dobaczewski1206, PhysRevC.90.034313} and references therein).

One question thus arises is how the density dependence of the zero range pairing interaction affects the pairing correlation. A systematic comparison between empirical OES from available experimental binding energies and BCS and HFB calculations with three different density dependent pairing forces has been done in Ref. \cite{PhysRevC.79.034306}. No significant difference was seen and it is suggested that there is a slight preference for the surface-peaked pairing \cite{PhysRevC.79.034306}. Such finding is consistent with the HFB calculations for the isotopic chain $^{100-132}$Sn \cite{PhysRevC.71.054303} and fission trajectories in superheavy nuclei \cite{doi:10.1142/S0218301307005740}. A mixed pairing force is used in the systematic study of Ref. \cite{nature486}.
On the other hand, in Ref.  \cite{Khan200794} it is shown that below the critical temperature where the pairing gap vanishes, the pairing gap is indeed sensitive to the surface or volume localization of the pairing force. 
Apparent differences were also noticed in the HFB calculations with the different density dependent pairing forces of neutron-rich Sn isotopes beyond $N=82$ in Refs. \cite{Dobaczewski2001361,Dobaczewski20021521,Dobaczewski01032002} and Ref. \cite{Duguet2005}. The effect of the density dependence of the pairing interaction on pairing
vibrations in $^{124,136}$Sn was analyzed with the HFB+QRPA approach in Ref. \cite{PhysRevC.80.044328}. The density dependence of the pairing may also influence the pair transfer properties of neutron Sn and light semi-magic neutron-rich nuclei \cite{PhysRevC.71.064306, PhysRevC.84.044317,PhysRevC.82.024318}.

This paper will examine systematically the effects of the density dependence of the pairing interaction on neutron-rich nuclei calculations within the HFB approach. The so-called volume, surface and mixed pairing force will be used. We will confront theoretical results with available experimental data and extend our calculations to the neutron drip line. We will show that, for neutron-rich nuclei, calculations with the surface pairing predict pairing gaps that are systematically stronger than those given by the mixed and volume pairing.
We will also investigate the neutron pairing correlation near the drip line from the view point of the di-neutron correlation. This work is partially motivated by a recent calculations presented in Ref. \cite{PhysRevC.88.034314} where HFB calculations with surface-peaked zero-range and finite-range pairing forces suggest that pairing can persist even in nuclei beyond the dripline.

The paper is organized as follows: In Sec. \ref{sec:MF}, we briefly discuss the HFB approach and the empirical OES from experimental binding energies. It is followed by the description of two-particle wave function. The HFB calculations with different pairing interactions are compared in  Sec. \ref{sec:Res}. A summary is given in Sec. \ref{sec:con}.

\section{The HFB approach and the pairing gap}
\label{sec:MF}

The HFB framework has been extensively discussed in the literature \cite{ring2004nuclear,Dobaczewski1984103,PhysRevC.53.2809,bender2003self,Dobaczewski1206}  and will only be briefly mentioned here for simplicity. In the standard HFB formalism, the Hamiltonian is reduced into two potentials, namely the mean field in the particle-hole channel and the pairing field in the particle-particle channel. It gives rise to the HFB equation 
\begin{equation}
\label{eq:HFB}
\begin{pmatrix}
 (H-\lambda) & \Delta \\
 -\Delta^{*} & -(H-\lambda)^{*}
 \end{pmatrix}
 \begin{pmatrix}
	 U_{k}\\
	 V_{k}
	\end{pmatrix}
        = E_{k} \cdot
	\begin{pmatrix}
	 U_{k}\\
	 V_{k}
	\end{pmatrix},  
\end{equation}
where $U_{k}$ and $V_{k}$ are the two components of single-quasi-particle wave functions.  In particle-hole channel we use the SLy4 Skyrme functional \cite{Chabanat1998}. In particle-particle channel we have the zero-range $\delta$  pairing force given as
\begin{equation}
\label{eq:pairing}
V_{pair}(\textbf{r},\textbf{r}^{\prime})= V_{0}\left(1-\eta \frac{\rho(\textbf{r})}{\rho_{0}}\right) \delta (\textbf{r}-\textbf{r}^{\prime}),
\end{equation}
where $V_{0}$ is the pairing strength, $\rho(\textbf{r})$ is the isoscalar local density and $\rho_{0}$ is the saturation density fixed at $0.16fm^{-3}$. $\eta$ takes the value $1$,$0$ and $1/2$ for surface, volume and mixed pairing, respectively. The pairing parameters are fitted to give a mean neutron gap of $1.31$MeV in ${}^{120}$Sn. The energy cutoff is $60$ MeV and the radius of the box is equal to $30$ fm.

In the present work we consider the HFB equation in spherical system in coordinate space with the Dirichlet boundary condition. The solutions are obtained with the HFB solver HFBRAD \cite{Bennaceur200596}. For comparison we also consider axially deformed solution of the Skyrme HFB equations in a harmonic oscillator basis using the HFBTHO code \cite{Stoitsov20131592}. 

We consider two different theoretical gaps: $\Delta_{LCS}$ canonical gap \cite{Lesinski2009}, which is the diagonal element of pairing-field matrix  for the Lowest Canonical State (LCS), and the \textit{average} gap $\Delta_{mean}$ that is the average values of the pairing fields \citep{Bennaceur200596}. These two theoretical pairing gaps were also compared with empirical pairing gaps recently in Ref. \cite{PhysRevC.88.034314}.

\subsection{Odd-Even mass difference}

The closest experimental data that we can compare our theoretical pairing gap with are the systematic variation of the nuclear binding energy depending on the evenness and oddness of number of proton $Z$ and neutron $N$. The OES effect has been extensively discussed in the literatures \cite{PhysRevLett.81.3599,PhysRevC.63.024308,PhysRevC.65.014311,bender2003self,PhysRevC.79.034306,
PhysRevC.80.027303}. The simplest form for OES is the three-point formula \cite{bohr1998nuclear,PhysRevLett.81.3599}, which has been extensively used for the empirical studies of the gap parameter $\Delta$.  For systems with even $N$ and fixed $Z$ the expression for the neutron pairing gap can be written as
\begin{eqnarray}
\label{eq:3pointC}
\Delta^ {(3)}_{C}(N)=\frac{1}{2}[S_n(N,Z)-S_n(N-1,Z)]\nonumber \\
=\frac{1}{2}\left[B(N,Z)+B(N-2,Z)-2B(N-1,Z)\right]
\end{eqnarray}
where $B$ is the (positive) binding energy which are extracted from Refs. \cite{ChinPhysC.36.1157,Nature.498.346349} and $S_n$ is the one-neutron separation energy. 
We will compare our results mainly with this three-point formula which actually corresponds to the conventional three-point formula for the case of odd
nuclei as \cite{Qi2012436,PhysRevC.83.014319,PhysRevC.79.034306},
\begin{multline}
\label{eq:3point}
\Delta^ {(3)}(N)=
-\frac{1}{2}\left[B(N-1,Z)\right.\\+\left. B(N+1,Z)-2B(N,Z)\right].
\end{multline} 
 There are other formulae such as the conventional three point\cite{bohr1998nuclear,PhysRevLett.81.3599}, four-point \cite{bohr1998nuclear,PhysRevLett.81.3599} and five-point \cite{NuclPhysA.476.1,NuclPhysA.536.20} formulae for calculating the pairing gap as
 \begin{multline}
\label{eq:4point}
\Delta^ {(4)}(N)= \frac{1}{4} [-B(N+1,Z)+3B(N,Z)\\
-3B(N-1,Z)+B(N-2,Z)]
\end{multline} 
and
\begin{multline}
\label{eq:5point}
\Delta^ {(5)}(N)= \frac{1}{8} [B(N+2,Z)-4B(N+1,Z)\\
+6B(N,Z)-4B(N-1,Z)+B(N-2,Z)].
\end{multline}
The direct comparison between the theoretical pairing gap and empirical OES is convenient from a computational point of view since only one single calculation is required and one avoids the complicated calculation of the odd nuclei. However, it should be mentioned that, even though they are quantitively quite close to each other in most cases, the theoretical gap is a model-dependent quantity and can not be compared with the empirical OES in a strict sense.

\subsection{Two-particle wave function}
In order to analyse  the clustering  feature  of two neutrons at the nuclear surface, we consider the spin-singlet component of two-particle wave function. The spatial structure of the  two-particle wave function can be written as (see, e.g., Ref. \cite{1742-6596-381-1-012131}),
\begin{eqnarray}
\Psi^{(2)} (r_1,r_2, \theta_{12})=\nonumber \\
 \frac{1}{4\pi}
\sum_{pq} \sqrt{\frac{2j_{p}+1}{2}} \delta_{l_pl_q}\delta_{j_pj_q}X_{pq}
 \phi_{p}(r_{1})\phi_{p}(r_{2})
P_{l_{p}}(\cos \theta_{12})
\end{eqnarray}
where $\phi$ is the single-particle wave function and $P_{l_{p}}$ is the Legendre polynomial. The two neutrons are at the distance $r_1$ and $r_2$ from the core, and $\theta_{12}$ is the angle between them. $X_{pq}$ is the expansion coefficient, which corresponds to the product $u_pv_q$ within the HFB approach. In this work, we obtain $\Psi^{(2)}$ as a function of $\theta_{12}$ and radius $r_{1}=r_{2}=R$.



\section{Results}
\label{sec:Res}

\subsection{Comparison between different OES formulae}

\begin{figure*}
\includegraphics[width=0.8\textwidth]{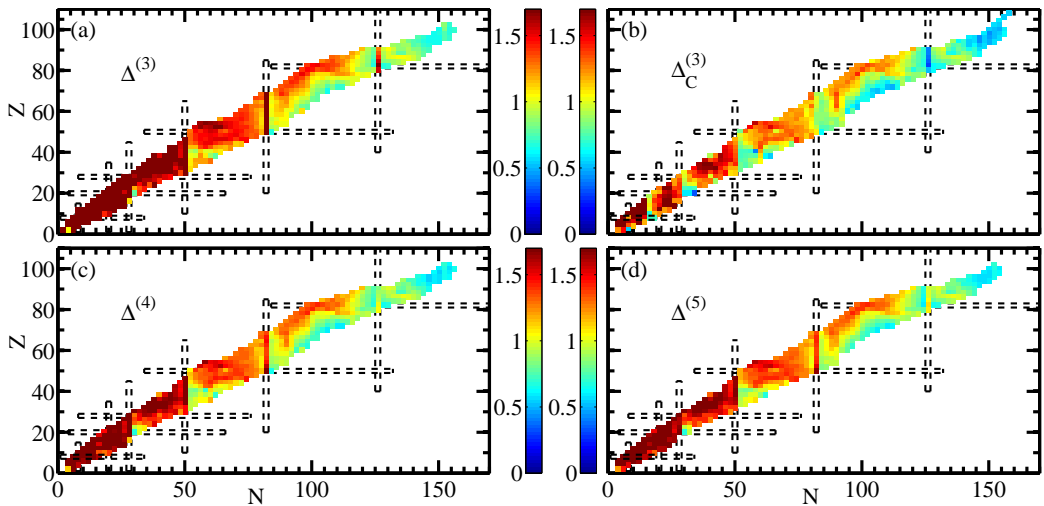}
\caption{\label{fig:4Gaps}  (Color online)  Neutron pairing gaps calculated for $\Delta^{(3)}$ (a, top left),  $\Delta^{(3)}_{C}$ (b, top right), $\Delta^{(4)}$ (c, bottom left) and $\Delta^{(5)}$ (d, bottom right) for all known even-even nuclei.
}
\end{figure*}

We begin our investigation by comparing the different OES formulae. In Fig. \ref{fig:4Gaps} we have plotted the results obtained from different OES formulae, namely 559 measured $\Delta^{(3)}$, 570 measured $\Delta^{(3)}_{C}$, 541 measured $\Delta^{(4)}$ and 516 measured $\Delta^{(5)}$ in even-even nuclei. For $\Delta^{(3)}$, almost all nuclei with $N<50$ have pairing gap larger than $1.7$ MeV. This is an indication of the large mean-field contribution in this region as mentioned in Ref. \cite{PhysRevLett.81.3599}. The shell effect for conventional OES-formula $\Delta^{(3)}$, $\Delta^{(4)}$ and $\Delta^{(5)}$ at neutron shell closure is also apparent in Fig. \ref{fig:4Gaps}.

\begin{figure}
\includegraphics[width=0.49\textwidth]{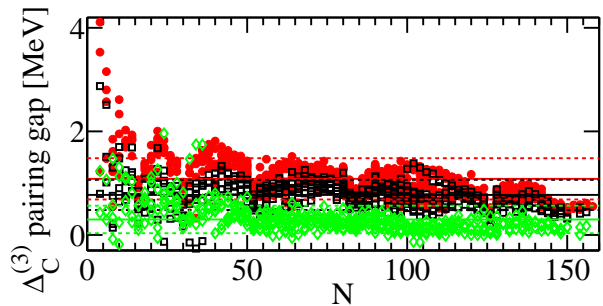}
\caption{\label{fig:oddevengap}  (Color online)  Neutron pairing gap $\Delta^{(3)}_{C}$ for even(odd) number of proton and even(even) number of neutron red circles (black squares). Green diamonds are proton-neutron interactions. Solid(dash) lines are the mean values($\pm \sigma$).  Gaps data with error more than $100$ keV are excluded.
}
\end{figure}

Fig. \ref{fig:oddevengap} shows the neutron $\Delta^{(3)}_{C}$ for even-even and even-odd nuclei. They show clearly the reduction of OES for even-odd number of nuclei by one rather constant magnitude of $\delta_{np}$ due to the extra binding in the intermediate odd-odd nuclei as a result of np correlation.
In Tab. \ref{tab:gap} the residual np interaction energy $\delta_{np}$ are obtained by reduction of pairings gaps of even-odd (even nuclei minus one) from even-even nuclei as
 \begin{multline}
\delta_{np}=\Delta^{(3)}_{C} (N,Z)- \Delta^{(3)}_{C} (N,Z-1)\\
=\frac{1}{2}[S_p(N,Z)+S_p(N-2,Z)]-S_p(N-1,Z).
\end{multline}
The case for odd-even nuclei can be defined in a similar way.
 The obvious trend as one may expect is that $\delta_{np}$ derived from proton gaps and neutron gaps are roughly the same and there is no visible dependence on shell closure.

\begin{table}
\caption{\label{tab:gap}Mean values (in MeV) of the residual proton-neutron interaction $\delta_{np}$ as extracted from the difference between neutron and proton pairing gaps $\Delta^{(3)}_{C}$ for even-even and those of the neighboring odd-$A$ nuclei. }
\begin{tabular}{c c c c c}
\hline
\hline
 & neutron & & proton & \\
\hline
$\delta_{pn}$  & $0.30\pm 0.26$ & & $0.31\pm 0.23$ & \\
\hline
\hline
\end{tabular}
\end{table}

We also evaluated the uncertainty of the extracted pairing gap in relation with 
the error in the experimental binding energy $\sigma_B(N,Z)$ by applying 
the error propagation as
\begin{eqnarray}
\begin{aligned}
\label{eq:sig}
\sigma_{\Delta}^{2}=\sum_{N,Z}\left(\frac{\partial \Delta}{\partial B(N,Z)}\right)^2{\sigma_{B(N,Z)}}^2,
\end{aligned}
\end{eqnarray}
where the sum runs over all nuclei involved in calculating the pairing gap $\Delta$.
The errors are quite small in most cases studied in this paper and remain invisible in the scales of our figures shown below.

\subsection{Systematic HFBTHO calculations for even-even nuclei}

 In order to explore the effects of the different pairing interactions on the pairing gap, we have firstly performed a global calculation using the HFBTHO code with the three different zero-range pairing interactions. A similar work was done in Ref. \cite{PhysRevC.79.034306} but only known nuclei were calculated. Our investigation is restricted to even-even nuclei for simplicity. All calculations are done in the usual harmonic oscillator basis by taking into account 25 major shells.

\begin{figure*}[htdp]
\includegraphics[width=0.7\textwidth]{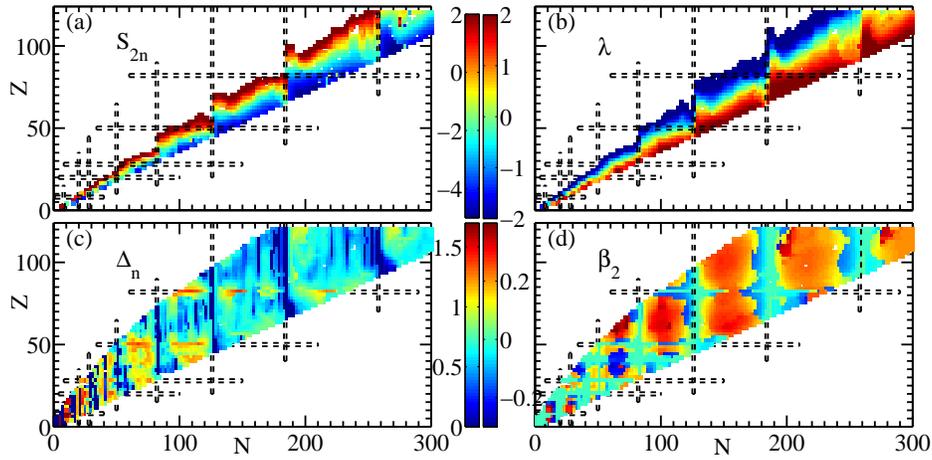}
\caption{\label{fig:MixHFB}  (Color online)  HFBTHO calculations with the mixed pairing interaction for the Fermi level $\lambda_{n}$ (b, top right panel), two neutron separation energy $S_{2n}=B(Z,N-2)-B(Z-N)$ (a, top left panel), mean neutron pairing gap $\Delta_{n}$ (c, bottom left panel) and the deformation $\beta$ (d, bottom right panel). }
\end{figure*}

\begin{figure*}[htdp]
\includegraphics[width=0.7\textwidth]{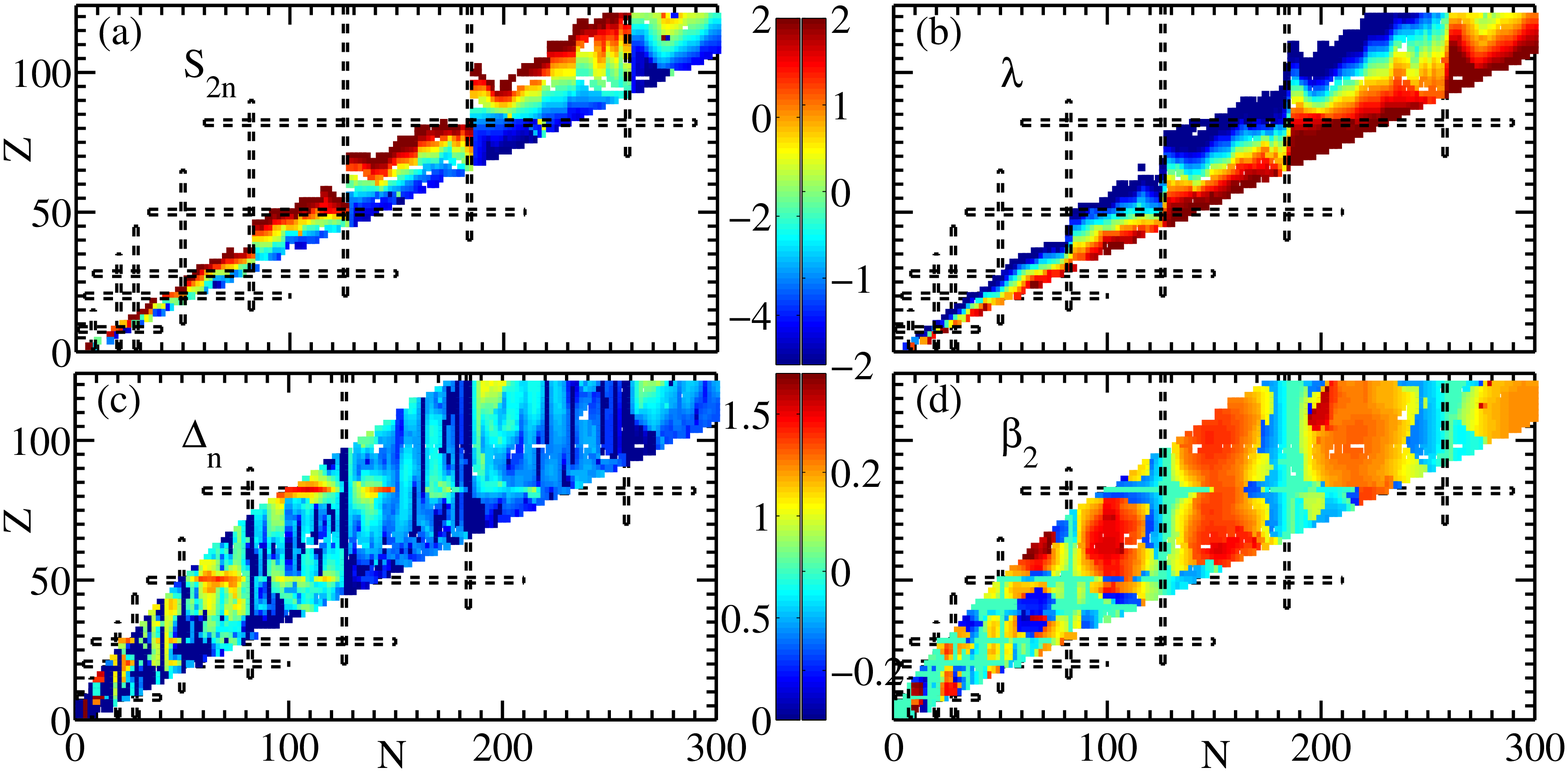}
\caption{\label{fig:VolHFB}  (Color online)  HFBTHO calculations with the volume pairing interaction for the Fermi level $\lambda_{n}$ (b, top right panel), two neutron separation energy $S_{2n}=B(Z,N-2)-B(Z-N)$ (a, top left panel), mean neutron pairing gap $\Delta_{n}$ (c, bottom left panel) and the deformation $\beta$ (d, bottom right panel).}
\end{figure*}

\begin{figure*}[htdp]
\includegraphics[width=0.7\textwidth]{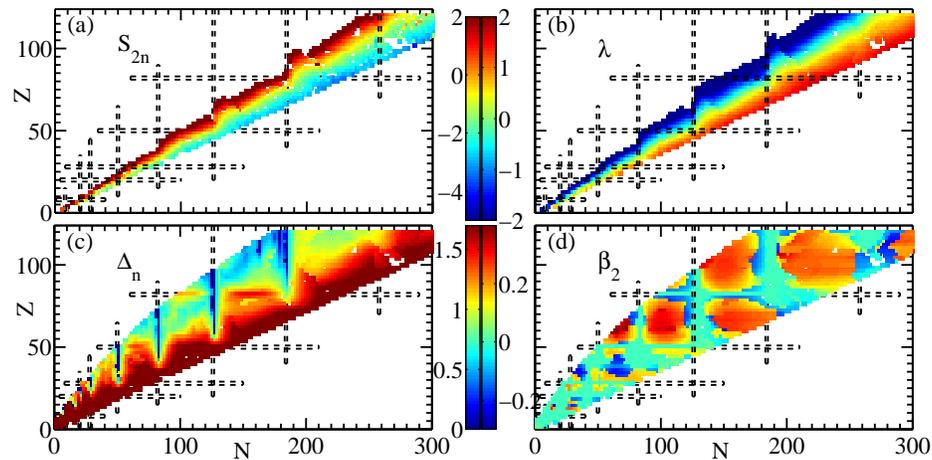}
\caption{\label{fig:SurfHFB}  (Color online)  HFBTHO calculations with the surface pairing interaction for the Fermi level $\lambda_{n}$ (b, top right panel), two neutron separation energy $S_{2n}$ (a, top left panel), mean neutron pairing gap $\Delta_{n}$ (c, bottom left panel) and the deformation $\beta$ (d, bottom right panel).}
\end{figure*}

Figs. \ref{fig:MixHFB}, \ref{fig:VolHFB} and \ref{fig:SurfHFB} show the Fermi level $\lambda_{n}$ , two neutron separation energy $S_{2n}$, mean neutron pairing gap $\Delta_{n}$ and the quadrupole deformation $\beta_2$ for mixed, volume and surface interactions, respectively. Only nuclei with Fermi level $\lambda_{n} > -3$ MeV and two neutron separation energy $S_{2n}<3$ MeV are included in the figures for a better comparison of nuclei around the neutron drip line.

\begin{figure*}[htdp]
\includegraphics[width=0.66\textwidth]{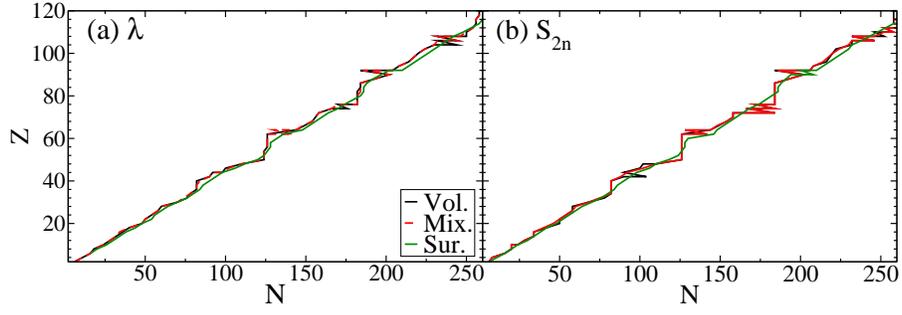}
\caption{\label{fig:drip} (Color online) The neutron driplines as defined by $\lambda=0$ (a, left) and $S_{2n}=0$ (b, right) given by the HFBTHO calculations with different pairing interactions.}
\end{figure*}

The major difference between these interactions is for nuclei close to drip line. It is found that calculations with the surface interaction predict a more smooth neutron dripline than the other interactions, as can be seen from Fig. \ref{fig:drip}. This is related to the fact that the pairing correlation in dripline nuclei predicted by calculations with the surface interaction is strong and overcomes the shell effect in many cases. 
Furthermore, by getting close to neutron drip line, mixed and volume interactions predict lower pairing gaps than those from the surface interaction.

Deformations calculated with the volume interaction are similar to those with the HFB approach with the Gogny force \cite{PhysRevC.81.014303}. Surface-interaction shows a different pattern for deformation for nuclei with $126<N<184$ and $N<50$.

\begin{figure*}[htdp]
\includegraphics[width=0.75\textwidth]{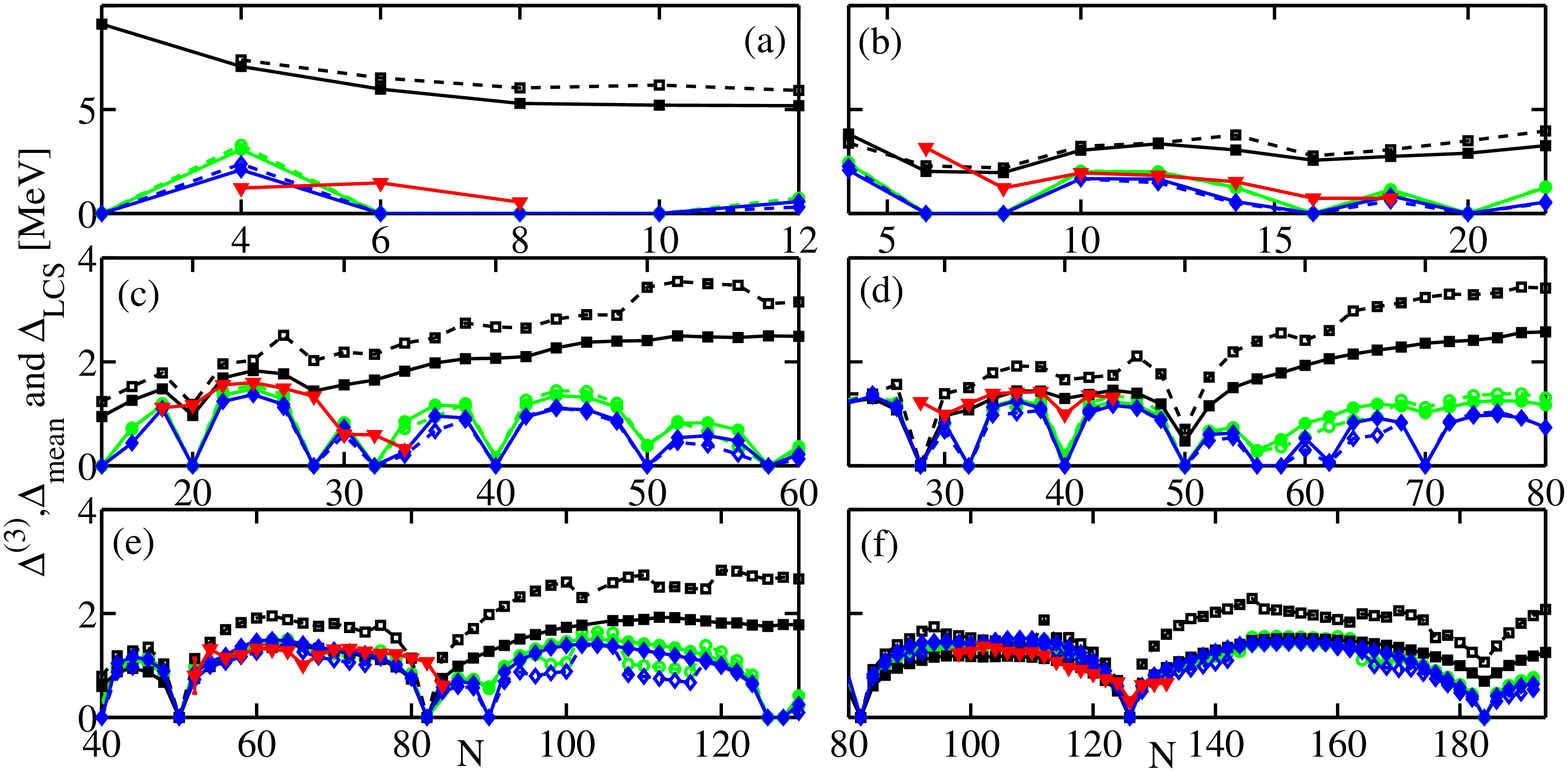}
\caption{\label{fig:hfbtho6isotopes}  (Color online)  HFBTHO calculations with mixed pairing (green circles), volume pairing (blue diamonds) and surface pairing (black squares) for the neutron pairing gaps in semimagic nuclei. The values of $\Delta_{LCS}$  (open markers) are connected by dashed lines while the average gaps ${\Delta}_{mean}$ (filled markers) are linked by solid lines. The red triangles correspond to the empirical pairing gaps $\Delta_C^{(N)}$.}
\end{figure*}

In Fig. \ref{fig:hfbtho6isotopes} we compare the two theoretical gaps, $\Delta_{LCS}$ and ${\Delta}_{mean}$, in semi-magic He, O, Ca, Ni, Sn and Pb isotopes calculated with the three different pairing forces. This may be compared to Figs. 2-3 in Ref. \cite{PhysRevC.88.034314}. It can be seen from the figure that the pairing gaps calculated from the surface pairing are systematically larger than those from the other two pairing forces in the light He and O isotopes and in neutron-rich nuclei shown in the figure. Moreover, there are noticeable differences between $\Delta_{LCS}$ and ${\Delta}_{mean}$ in the surface pairing calculations whereas those two values are pretty close to each other in the other calculations with the mixed and volume pairing forces. The pairing gaps predicted by the mixed and volume pairing forces are similar in most cases.

Moreover, as can be seen from Figs. \ref{fig:SurfHFB} and \ref{fig:hfbtho6isotopes}, calculations with the surface pairing interaction predict large pairing gaps for neutron rich nuclei both around and beyond the dripline. The pairing correlation in nuclei in the neutron-rich region given by this calculation can be significantly stronger than those of the stable nuclei and can even overcome the shell effect in many cases.

\subsection{HFBRAD calculations for semi-magic even-even nuclei}

\begin{figure*}[htdp]
\includegraphics[width=0.75\textwidth]{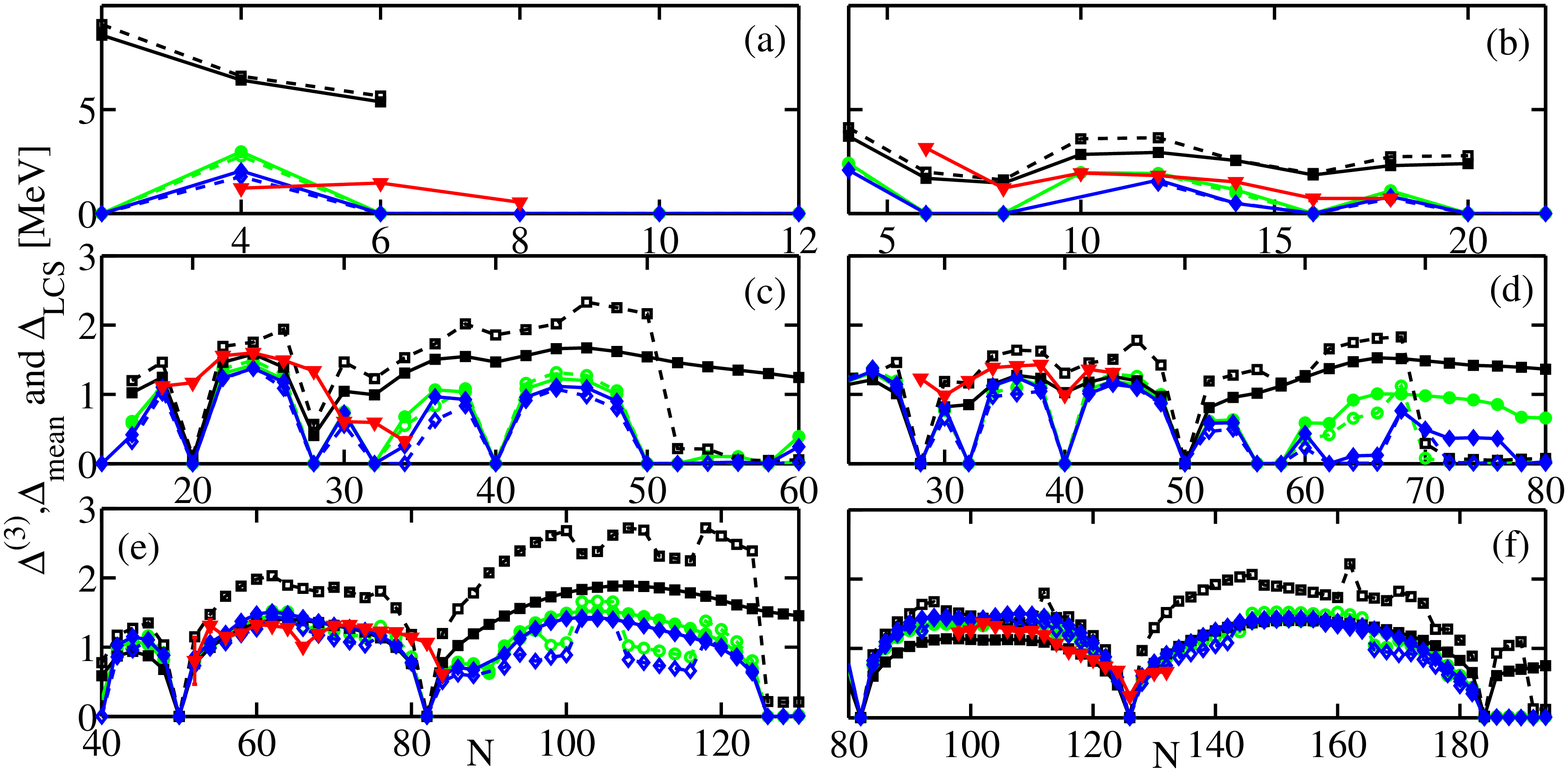}
\caption{\label{fig:6isotopes}  (Color online)  HFBRAD calculations with mixed pairing (green circles), volume pairing (blue diamonds) and surface pairing (black squares) for the neutron pairing gaps in semi-magic nuclei. The values of $\Delta_{LCS}$  (open markers) are connected by dashed lines while the average gaps ${\Delta}_{mean}$ (filled markers) are linked by solid lines. The red triangles correspond to the empirical pairing gaps $\Delta_C^{(N)}$.}
\end{figure*}

It is expected that calculations in the coordinate space may provide a more precise description for weakly bound nuclei in the vicinity of the dripline. Thus in Fig. \ref{fig:6isotopes} we have redone the calculations presented in Fig. \ref{fig:hfbtho6isotopes}  with the HFBRAD code. All calculations presented in the figures are done by restricting the maximal spin to be $j=25/2$ except the light He, O, Ca and Ni isotopes where we take $j\leq$ 9/2, 11/2, 13/2 and 15/2, respectively. We have also done calculations for those nuclei by extending the spin up to j=25/2. However, as we will also mention below, the pairing gaps thus calculated will be significantly overestimated if the surface pairing is used.
Fig. \ref{fig:6isotopes} shows clearly again that volume and mixed pairing can reproduce well the magnitude of the observed $\Delta^{(3)}_{C}$ for both the light and heavier semi-magic nuclei. However, there is no consistency  in case of surface interaction. As can be seen from the figure, for calcium, nickel, tin and lead isotopes, all three pairing interactions agree well with the experimental data in most cases. Significant differences between predictions of the surface interaction and those of the mixed and volume interactions are seen in unknown regions with no experimental data as well as in light He and O isotopes. Calculations with the surface interaction are also much more sensitive to the number of shells considered than those of the mixed and volume pairing calculations. This is also related to the fact that the pairing matrix elements predicted by the surface pairing are much larger than those of the mixed and volume pairing for weakly bound and unbound levels.

As can be seen from Figs. \ref{fig:hfbtho6isotopes} \& \ref{fig:6isotopes}, both calculations in the HO and coordinate spaces with the surface interaction predict large neutron pairing gaps for nuclei on the neutron-rich side. A noticeable difference between the two calculation is that, in the latter case, the calculated $\Delta_{LCS}$  vanish for Ca, Ni, Sn and Pb isotopes beyond the dripline whereas the mean gaps persist in some cases. This has also been noticed in Ref. \cite{PhysRevC.88.034314}. The theoretical $\Delta_{LCS}$ and ${\Delta}_{mean}$ values are quite close to each other in most cases in both calculations with the mixed and volume pairing forces. They drop to zero when one goes beyond neutron dripline for all semi-magic nuclei studied here except Ni isotopes.

\subsection{Di-neutron correlation in neutron-rich Ni isotopes}

\begin{figure}
\includegraphics[width=0.46\textwidth]{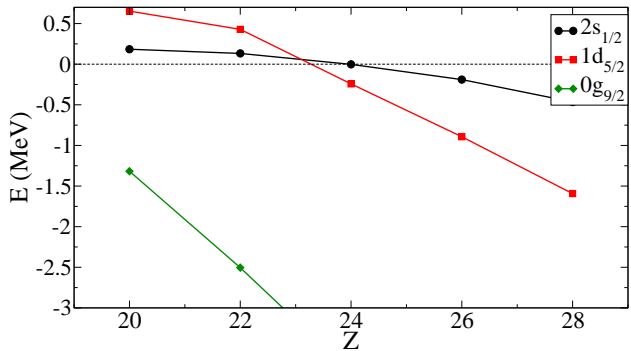}
\caption{\label{figspe}  (Color online)  SHF calculations with the Sly4 force on the evolution of the single-particle energies in the neutron-rich $N=52$ isotones. With no pairing considered, the spurious $s_{1/2}$ states with positive energies have no physical meaning and are shown only to illustrate the tendency.}
\end{figure}

Nuclei around the neutron-rich isotope $^{78}$Ni, which may become accessible experimentally soon, are of particular interest in relation to the search for the loosely bound $2s_{1/2}$ orbital and neutron halo that may thus form. The $s_{1/2}$ neutron orbital near threshold show a behavior that is quite different from other orbitals with larger orbital angular momentum: They lose energy in a way that is much slower than other orbitals when the potential becomes shallower (see, e.g., Refs. \cite{PhysRevC.89.061305, Xu2013247} and references therein). As an example, in Fig. \ref{figspe} we plot the the evolution of the single-particle energies in the neutron-rich $N=52$ isotones. As can be seen from the figure, as one removes protons and the mean field gets shallower, the $1d_{5/2}$ and $0g_{9/2}$ neutron orbitals lose their energies much faster than that of $2s_{1/2}$. One may expect that a loosely bound $s_{1/2}$ may be found in this region below the $d_{5/2}$ and $g_{7/2}$ orbitals. The situation may be further perturbed by considering the pairing effect.

\begin{figure}
\includegraphics[width=0.4\textwidth]{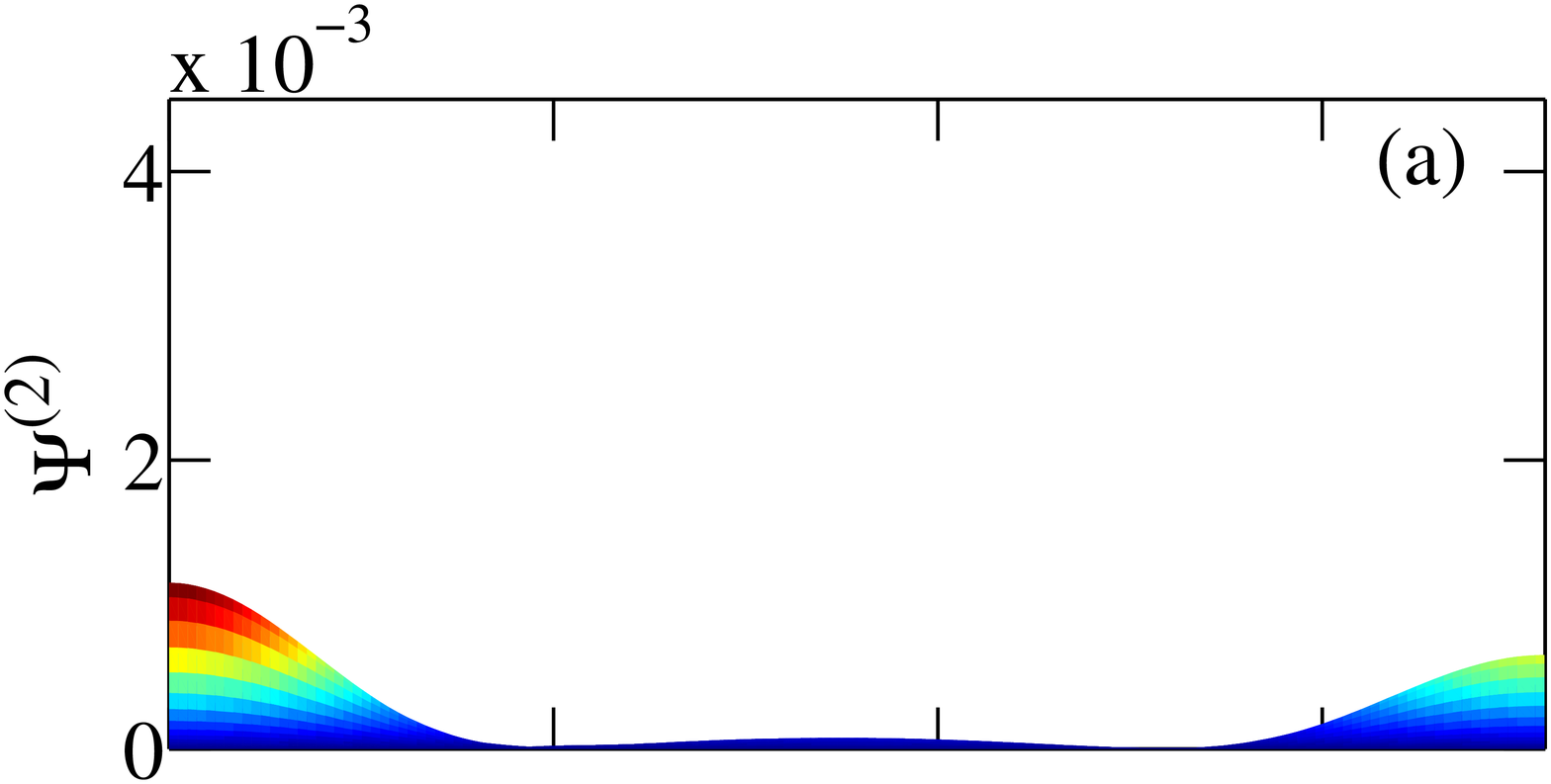}
\includegraphics[width=0.4\textwidth]{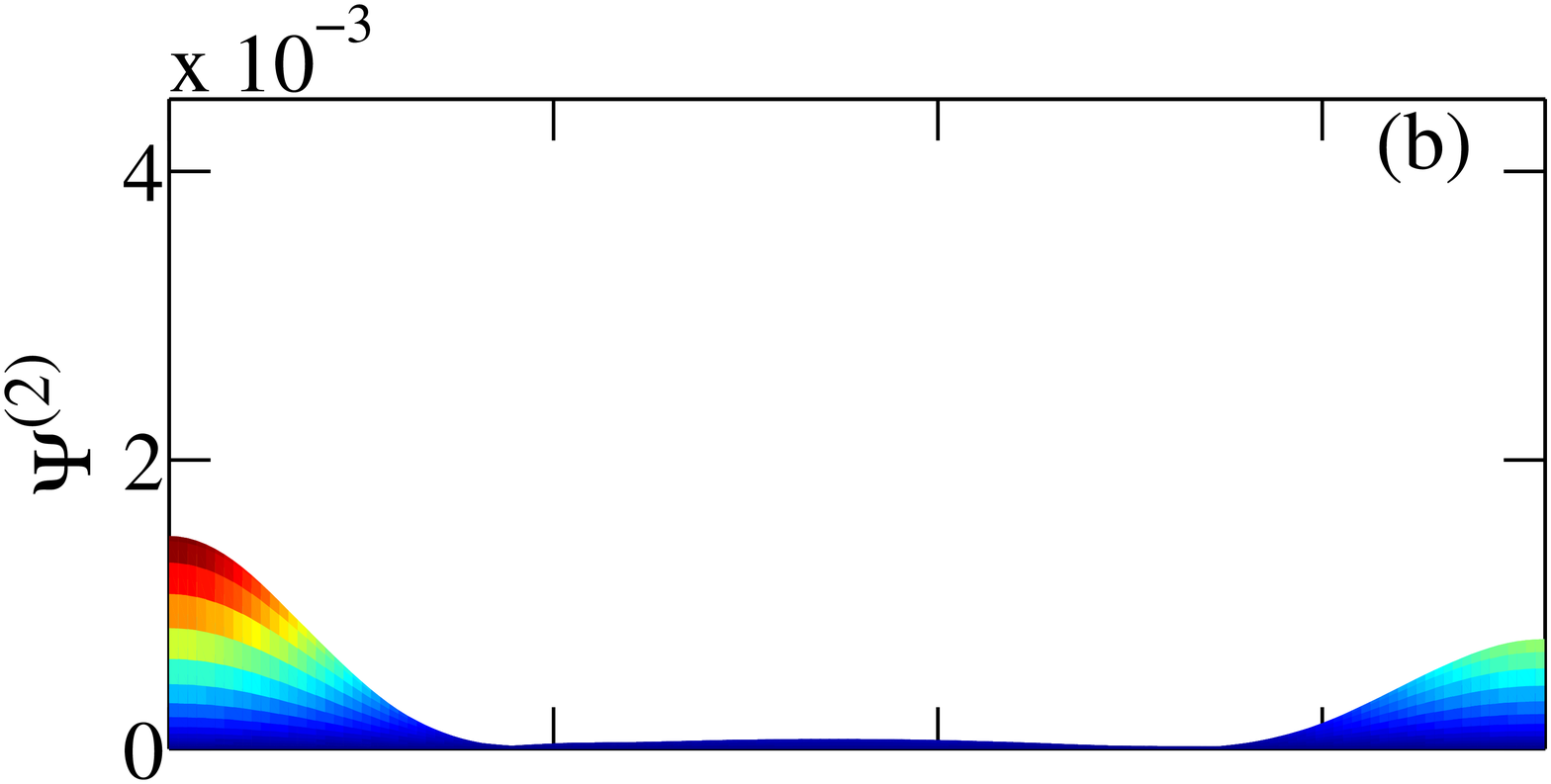}
\includegraphics[width=0.4\textwidth]{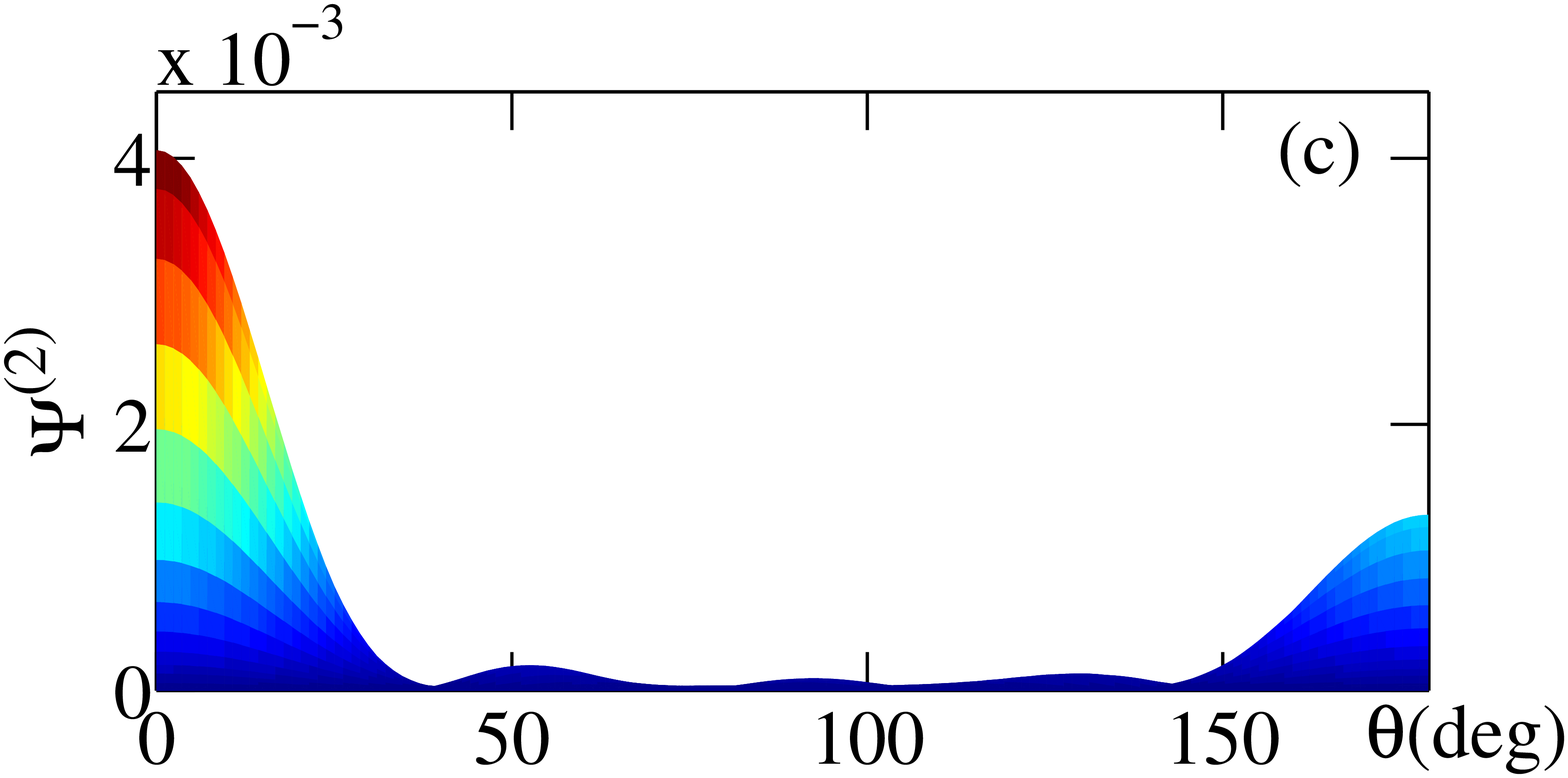}
\caption{\label{fig:twoparticleNi}  (Color online)  Two-particle wave function $\Psi^{(2)}$ for $^{82}$Ni. Calculations with the volume, mixed and surface interactions are shown at  the top, middle and bottom row, respectively. }
\end{figure}

\begin{table*}[htdp]
\caption{\label{tab:Ni8082} Calculations with different pairing forces on the chemical potential $\lambda_{n}$, pairing gaps, and the occupancy of the $2s_{1/2}$ neutron orbital in neutron-rich $^{82-88}$Ni isotopes.}

\begin{tabular}{l | c c c c | c c c c | c c c c | c c c c}
\hline
\hline
isotopes & & ${}^{82}$Ni & & & & ${}^{84}$Ni &&& &  ${}^{86}$Ni && & & ${}^{88}$Ni  \\
\hline
Interaction & $\lambda_{n}$ & $\Delta_{mean}$& $\Delta_{LCS}$ & $v^{2}$ & $\lambda_{n}$& $\Delta_{mean}$& $\Delta_{LCS}$ & $v^{2}$ & $\lambda_{n}$& $\Delta_{mean}$& $\Delta_{LCS}$  & $v^{2}$  & $\lambda_{n}$& $\Delta_{mean}$& $\Delta_{LCS}$ & $v^{2}$ \\
\hline
$j_{max}=15/2$\\
Volume  & $-1.68$ & $0.59$ & $0.50$ & $2\%$ & $-1.21$ & $0.0$ & $0.0$& $0\%$& $-0.57$ & $0.0$ & $0.0$& $100\%$& $-0.15$ & $0.43$ & $0.23$ & $99\%$ \\

Mixed & $-1.66$  & $0.63$ & $0.58$& $3\%$ & $-1.22$ & $0.0$& $0.0$& $0\%$ &$-0.58$ & $0.0$& $0.0$& $100\%$ &$-0.17$ & $0.59$& $0.35$& $97\%$ \\

Surface  & $-1.66$ &$0.96$& $1.28$ & $11\%$  & $-1.25$ &$1.01$& $1.36$& $25\%$ &$-0.82$ & $1.12$ & $1.16$ & $52\%$ & $-0.45$ & $1.25$ & $1.27$ & $72\%$\\
\hline
$j_{max}=25/2$\\
Volume  & $-1.67$ & $0.62$ & $0.53$ & $2\%$ & $-1.21$ & $0.0$ & $0.0$& $0\%$& $-0.57$ & $0.0$ & $0.0$& $100\%$& $-0.16$ & $0.48$ & $0.25$ & $98\%$ \\

Mixed & $-1.64$  & $0.72$ & $0.66$& $4\%$ & $-1.22$ & $0.27$& $0.25$& $4\%$ &$-0.60$ & $0.35$& $0.22$& $92\%$ &$-0.20$ & $0.75$& $0.46$& $94\%$ \\

Surface  & $-1.93$ &$1.46$& $2.12$ & $15\%$  & $-1.50$ &$1.61$& $2.29$& $27\%$ &$-1.10$ & $1.74$ & $2.48$ & $41\%$ & $-0.74$ & $1.87$ & $2.28$ & $54\%$\\

\hline
\hline
\end{tabular}
\end{table*}

\begin{table*}[htdp]
\caption{\label{tab:Nistr} Same as Table \ref{tab:Ni8082} but for calculations with the strengths of pairing interaction enhanced by 5\%.}

\begin{tabular}{l | c c c c | c c c c | c c c c | c c c c}
\hline
\hline
isotopes & & ${}^{82}$Ni & & & & ${}^{84}$Ni &&& &  ${}^{86}$Ni && & & ${}^{88}$Ni  \\
\hline
Interaction & $\lambda_{n}$ & $\Delta_{mean}$& $\Delta_{LCS}$ & $v^{2}$ & $\lambda_{n}$& $\Delta_{mean}$& $\Delta_{LCS}$ & $v^{2}$ & $\lambda_{n}$& $\Delta_{mean}$& $\Delta_{LCS}$  & $v^{2}$  & $\lambda_{n}$& $\Delta_{mean}$& $\Delta_{LCS}$ & $v^{2}$ \\
\hline
$j_{max}=15/2$ $5\%$\\
Volume   & $-1.66$ & $0.69$  & $0.58$  & $3\%$  & $-1.21$ & $0.0$  & $0.0$ & $0\%$ & $-0.57$ & $0.0$   & $0.0$   & $100\%$ & $-0.16$ & $0.55$ & $0.28$  & $97\%$ \\

Mixed   & $-1.63$  & $0.76$ & $0.69$& $4\%$ & $-1.21$ & $0.32$& $0.28$& $5\%$ &$-0.60$ & $0.36$  & $0.22$ & $92\%$ &$-0.20$ & $0.78$& $0.46$& $94\%$ \\

Surface & $-1.73$ &$1.21$& $1.76$ & $13\%$  & $-1.31$ &$1.31$& $1.74$& $26\%$ &$-0.90$ & $1.42$  & $1.89$ & $47\%$ & $-0.53$ & $1.54$ & $1.59$ & $65\%$\\
\hline
$j_{max}=25/2$ $5\%$\\
Volume  & $-1.65$ & $0.74$ & $0.62$ & $3\%$ & $-1.21$ & $0.0$ & $0.0$& $0\%$& $-0.57$ & $0.0$ & $0.0$& $100\%$& $-0.16$ & $0.62$ & $0.32$ & $98\%$ \\

Mixed & $-1.62$  & $0.86$ & $0.81$& $5\%$ & $-1.19$ & $0.67$& $0.61$& $14\%$ &$-0.64$ & $0.74$& $0.48$& $76\%$ &$-0.26$ & $1.0$& $0.62$& $90\%$ \\

Surface  & $-2.12$ &$1.84$& $2.66$ & $17\%$  & $-1.67$ &$1.99$& $2.84$& $27\%$ &$-1.27$ & $2.13$ & $3.03$ & $38\%$ & $-0.91$ & $2.25$ & $3.21$ & $49\%$\\

\hline
\hline
\end{tabular}
\end{table*}

To further analyze the effect of the pairing, as a typical example, in Fig.  \ref{fig:twoparticleNi} we show the square of two-neutron wave function $|\Psi^{2\nu}(r,r,\theta)|^2$  for the nucleus $^{82}$Ni calculated with the HFBRAD code with different paring interactions. 
In the figure $|\Psi^{2\nu}(r,r,\theta)|^2$ are calculated in a mesh defined by $r$ and $\theta$ but then projected on a two-dimensional plane for a clearer vision. In this way one can make sure that the peaks shown corresponds to the real ones. Those peaks appear around $r=5.2$ fm in all three cases.
As can be seen from the figure, the di-neutron correlation predicted by the surface pairing interaction calculation is much stronger than those from the mixed and volume pairing interactions. This is related to that fact that calculations with the surface pairing give much larger pairing gaps than the other calculations. As a result, one needs a significantly larger model space to get convergence in that calculation and big differences are seen between calculations with maximal spin values $j=15/2$ and $25/2$. The wave functions derived from surface pairing calculations are also significantly more mixed. In Table \ref{tab:Ni8082}  we give the calculated chemical potentials $\lambda_{n}$, pairing gaps, and the occupancies of the $2s_{1/2}$ neutron orbital in neutron-rich $^{82-88}$Ni isotopes with the three different pairing forces. Calculations with the surface pairing predict a significant mixture between the $s_{1/2}$ orbital and neighboring ones.
The surface pairing calculation also predicts a deeper chemical potential and larger pairing gaps than the other two calculations. Moreover, as can seen from Table \ref{tab:Nistr}, calculations with the surface pairing are much more sensitive to the strength of the pairing than those of the other two pairing interactions.

\section{Summary} 
\label{sec:con}

In this work we present a systematic study on the neutron pairing gaps predicted by HFB calculations with the Skyrme force and zero-range pairing forces with different density dependence. We first compared the experimental pairing gaps from four different OES formulae.
 Then we applied the HFB approach to study the pairing correlations in even-even nuclei including the neutron-rich semi-magic even-even nuclei. We tested the different volume, mixed and surface pairing interactions with the  SLy4 parameterization of the Skyrme interaction in the particle-hole channel.
 
 It is found that different treatments of pairing force can affect the calculated $\Delta_{LCS}$ and $\Delta_{mean}$ significantly in neutron-rich nuclei in the vicinity of drip line. Whereas the effect is much less visible in calculations for known nuclei. Moreover, our calculations show that the pairing gaps given by the surface-peaked pairing interaction are systematically larger than those of the volume and mixed pairing forces. 
Beyond the neutron dripline, there is a clear difference between mean gap and lowest canonical gap in calculations in coordinate representation with the surface-pairing interaction. This is not seen in calculations with other pairing forces.  Moreover, the di-neutron correlations in unstable nuclei and the position of the two-neutron dripline can be quite different depending on the density dependence of the pairing force.

\section*{Acknowledgement}
This work was supported by the Swedish Research Council (VR) under grant Nos. 621-2012-3805, and 621-2013-4323. The calculations were performed on resources provided by the Swedish National Infrastructure for Computing (SNIC) at NSC in Link\"oping.

\end{document}